\documentclass[aps,pre,onecolumn,showpacs,floatfix]{revtex4}
\usepackage{amssymb,amsmath,epsfig,graphicx}

\usepackage{caption}
\usepackage{dcolumn}

\newcommand{\be}{\begin{equation}}
\newcommand{\ee}{\end{equation}}
\newcommand{\bea}{\begin{eqnarray}}
\newcommand{\eea}{\end{eqnarray}}
\newcommand{\bi}{\begin{itemize}}
\newcommand{\ei}{\end{itemize}}

\newcommand{\uoff}{u_{\text{off}}}
\newcommand{\uon}{u_{\text{{on}}}}
\newcommand{\fon}{f_{\text{on}}}
\newcommand{\foff}{f_{\text{off}}}
\newcommand{\vmedia}{\langle \dot x_{\text{cm}}\rangle_{\text{st}}}
\newcommand{\xmedia}{ \Delta x_{\text{cm}} }
\newcommand{\xcm}{x_{\text{cm}}}

\newcommand{\calTon}{{\cal T}_{\text{on}}}
\newcommand{\calToff}{{\cal T}_{\text{off}}}
\newcommand{\dfdt}{\dot f_{\text{{exp}}}}

\newcommand{\ton}{t_{\text{on}}}
\newcommand{\toff}{t_{\text{off}}}
\newcommand{\xon}{x_{\text{on}}}
\newcommand{\Con}{C_{\text{on}}}
\newcommand{\Coff}{C_{\text{off}}}
\newcommand{\lambdaon}{\lambda_{\text{on}}}
\newcommand{\lambdaoff}{\lambda_{\text{off}}}
\newcommand{\tauon}{\tau_{\text{on}}}
\newcommand{\tauoff}{\tau_{\text{off}}}

\begin{document}

\title{\bf Threshold feedback control for a collective flashing ratchet: threshold dependence}

\author{M. Feito}
\email{feito@fis.ucm.es}
\affiliation{Departamento de F\'{\i}sica At\'omica, Molecular y Nuclear,
Universidad Complutense de Madrid, \\
Avenida Complutense s/n, 28040 Madrid, Spain}

\author{F. J. Cao}
\email{francao@fis.ucm.es}
\affiliation{Departamento de F\'{\i}sica At\'omica, Molecular y Nuclear,
Universidad Complutense de Madrid, \\
Avenida Complutense s/n, 28040 Madrid, Spain.}
\affiliation{LERMA, Observatoire de Paris, Laboratoire Associ\'e au CNRS UMR 8112, \\
61, Avenue de l'Observatoire, 75014 Paris, France.}

\date{\today}

\begin{abstract}
We study the threshold control protocol for a collective flashing
ratchet. In particular, we analyze the dependence of the current
on the values of the thresholds. We have found analytical
expressions for the small threshold dependence both for the few
and for the many particle case. For few particles the current is a
decreasing function of the thresholds, thus, the maximum current
is reached for zero thresholds. In contrast, for many particles
the optimal thresholds have a nonzero finite value. We have
numerically checked the relation that allows to obtain the optimal
thresholds for an infinite number of particles from the optimal
period of the periodic protocol. These optimal thresholds for an
infinite number of particles give good results for many particles.
In addition, they also give good results for few particles due to
the smooth dependence of the current up to these threshold values.
\end{abstract}

\maketitle


\section{Introduction} \label{sec:intro}

Ratchets or Brownian motors are rectifiers of thermal
fluctuations. This rectification is usually achieved through the
introduction of an external deterministic or stochastic
perturbation in a system that is or becomes asymmetric under
spatial inversion~\cite{revreim}. Over the last years ratchets
have been studied due to their theoretical and experimental
relevance. From a practical point of view the ratchet effect has
many potential applications in biology, condensed matter and
nanotechnology~\cite{revreim,lin02}.

Ratchets can be viewed as controllers that act on stochastic
systems with the aim of inducing directed motion through the
rectification of the fluctuations. In particular, flashing
ratchets are thermal fluctuation rectifiers based on switching on
and off a periodic potential \cite{pro92,ast94}. Several studies
deal with the problem of optimizing the particle current
\cite{optflux} or the efficiency \cite{opteff} in these systems.
However, they all consider only open-loop controllers (as that
obtained with a periodic or random switching). Recently, feedback
control protocols have been introduced in the context of
collective ratchets \cite{prl,europhys}. In the feedback control
protocols the action of the controller depends on the state of the
system. This feedback control, or closed-loop control, can be
implemented in systems where the state of the system is monitored
(as occurs in some experimental setups with colloidal particles
\cite{exp}).

In this paper we study one of these closed-loop controls, the
\emph{threshold control}, previously introduced in Ref.~\cite{europhys}.
The structure of the paper is as follows. In the next Section we
present the mathematical model of the collective flashing ratchet
with the threshold control protocol and we discuss briefly other
protocols that have been studied in recent articles. Later, in
Sec.~\ref{threshold}, we analyze the dependence of the average
center-of-mass velocity on the thresholds, obtaining analytical
approximated expressions that are compared with the numerical
results. In Subsec.~\ref{small} we study the small thresholds
dependence (distinguishing the many particles case and the few
particles case), while in Subsec.~\ref{general} we discuss the
dependence of the average center-of-mass velocity for any
thresholds and any number of particles. Finally, we present our
conclusions in Sec.~\ref{conclusions}.

\section{The model}

We consider $N$ Brownian particles with positions $x_i(t)$ at
temperature $T$ within a ratchet potential $V(x)$, and whose
dynamics is described by the overdamped Langevin equations
\begin{equation}\label{langevin}
\gamma \dot x_i(t)=\alpha(t)F(x_i(t))+\xi_i(t);\qquad i=1,\dots,N,
\end{equation}
with $\gamma$ the friction coefficient (related to the diffusion coefficient
$D$ through Einstein's relation $D=k_BT/\gamma$) and $\xi_i(t)$ Gaussian
white noises of zero mean satisfying the fluctuation-dissipation relation
$\langle
\xi_i(t)\xi_j(t^\prime)\rangle =2\gamma k_B
T\delta_{ij}\delta(t-t^\prime)$. The force is given by
$F(x)=-V^\prime (x)$ and $\alpha$ is a control parameter that can
take only two possible values, $\alpha=0$ (potential `off') or
$\alpha=1$ (potential `on').

Several control strategies have been studied in order to maximize
the particle current in this system. The
\emph{optimal periodic switching}~\cite{europhys,prl} consists on
switching the potential on during a time period $\calTon$ and
switching it off during $\calToff$. Note that it is an open-loop
control protocol and therefore the results are independent of $N$.
This protocol is the periodic flashing ratchet, that has been
widely studied both theoretically and experimentally
\cite{revreim,lin02}. The
\emph{maximization of the center-of-mass instant velocity} protocol
has been introduced and studied in Ref.~\cite{prl}. It consist on
switching the potential on only if the net force would be
positive. Therefore, it is a closed-loop control protocol, because it needs
information about the state of the system in order to operate.
This is the best strategy for a single particle.
However, for a large number of particles the system gets trapped
with the potential `on' or `off' and then the average steady state
current tends to zero as $N$ increases~\cite{prl}. Another closed-loop control
protocol, the \emph{threshold control}, was proposed in~\cite{europhys} to
avoid this effect. In this paper we analyze it further.
\par

The net force per particle is
\begin{equation}
f(t)=\frac{1}{N}\sum_{i=1}^N F(x_i(t)).
\end{equation}
On the other hand, given the state of the system $ x_i(t) $,
a good estimator for the time derivative of $ f(t) $ can be obtained
using Langevin equation \eqref{langevin} and
Ito calculus (see Ref.~\cite{europhys}),
\begin{equation}
\dfdt \equiv \frac{1}{\gamma N}\sum_i
\alpha(t)F(x_i(t))F^\prime(x_i(t)) +\frac{k_B T}{\gamma N}\sum_i
F^{\prime\prime}(x_i(t)) .
\end{equation}

The \emph{maximization of center-of-mass instant velocity} protocol has
$\alpha(t)=\Theta(f(t))$, with $\Theta$ the Heaviside
function [$\Theta (x)=1$ if $x>0$, else $\Theta (x)=0$].
In contrast, the \emph{threshold control} policy has two thresholds
$ \uon \geq 0 $ and $ \uoff \leq 0 $ which induce earlier switchings
that permit to avoid the trapping. When $f(t)$ decreases below $\uon$
we switch off the potential, although the net force is still positive,
in order to avoid the trapping. Analogously the potential is switched on if
the net force per particle increases above $\uoff$, so we induce
the flipping of the system before $f(t)$ is positive. Therefore,
the \emph{threshold control} is given by
\begin{equation} \label{alfa}
\alpha(t)=
\begin{cases}
1& \mbox{if $f(t)\geq \uon$,}\\
1& \mbox{if $\uoff<f(t)<\uon$ and $\dfdt(t)\geq 0$,}\\
0& \mbox{if $\uoff<f(t)<\uon$ and $\dfdt(t)< 0$,}\\
0& \mbox{if $f(t)\leq \uoff$.}
\end{cases}
\end{equation}
This scheme removes the long decaying tails in the evolution of
the net force preventing the trapping. Note that this protocol and
the maximization of the center-of-mass instant velocity
protocol are feedback controls or closed-loop controls.
The threshold control protocol in the zero thresholds limit
gives the maximization of the center-of-mass instant velocity
protocol.

\section{Threshold control strategy}\label{threshold}

\subsection{Small thresholds}\label{small}

In this subsection we analyze the threshold control strategy
improving and extending the analytic expressions found for the maximization of
the center-of-mass instant velocity protocol \cite{prl}.

\subsubsection{Many particles: quasideterministic approximation}\label{many}

For many particles (large $N$) the net force has a
quasideterministic behavior. It can be described in
terms of two contributions, a deterministic contribution $ f^\infty $
(given by the behavior for an infinite number of particles) plus a
small stochastic contribution
\begin{equation}\label{cuasideterminista}
f(t)=f^\infty (t)+\mbox{fluctuations}.
\end{equation}
This approximate description has proven to be fruitful in order to
understand the behavior of these ratchets in the many particle
case~\cite{prl}.

The deterministic contribution, that reflects the behavior of the system
for an infinite number of particles ($N\to \infty$), can be
described through a particle distribution $\rho(x,t)$ that evolves
according to the mean-field Fokker-Planck equation
$ \gamma \partial_t \rho(x,t)=\left[ -\alpha (t)\partial_x
  F(x)+k_BT\partial^2_x\right]\rho(x,t) $.
The net force per particle is a deterministic function $
f^{\infty}(t)= \langle F(x) \rangle_\rho \equiv \int_0^L dx \,\rho
(x,t) F(x) $, with $ L $ the period of the ratchet potential.
The net force is zero for the equilibrium
distribution when the potential is on and also when it is off. We
denote by $ f_\nu^\infty (t) $ with $\nu=\mbox{on,off}$ the value
of the deterministic part of the net force when the system has
been evolving with the potential on or off respectively a time $ t
$ after a switching. After a certain time $ \tau_{\nu} $ it can be
approximately described by \cite{prl}
\begin{equation}\label{exponencial}
f_\nu^\infty (t) = C_\nu e^{-\lambda_{\nu} (t-\tau_\nu)}.
\end{equation}
$ C_\nu $ and $ \lambda_\nu $ are constants that are obtained by
fitting the net force obtained with the Fokker-Planck equation. In
order to obtain $ \fon^\infty(t) $ we evolve the equilibrium
distribution for the off potential with the Fokker-Planck equation
with the potential on, i.e., we assume that the system was close
to the equilibrium state for the off potential before the off-on
switching. We proceed analogously for $\foff^\infty(t)$.

On the other hand, the amplitude of the fluctuations of the net force $ f $
can be estimated as \cite{prl}
\begin{equation} \label{Sigma}
\Sigma = \sqrt{ \langle f^2(t) \rangle - \langle
f(t) \rangle^2 }
  \simeq \sqrt{ \frac{ \langle F^2 \rangle_\rho - \langle F \rangle_\rho^2
    }{N}}
  \sim \frac{V_0}{L\sqrt{a(1-a)N}}
  .
\end{equation}
This simple result is a good estimation of the amplitude of the
fluctuations for potentials with characteristic height $ V_0 $
and asymmetry $ a $. For example, the potential
\begin{equation} \label{smoothpot}
V(x)=\frac{2V_0}{3\sqrt{3}}\left[ \sin\left(\frac{2\pi
x}{L}\right)+\frac{1}{2} \sin\left(\frac{4\pi x}{L}\right)\right],
\end{equation}
that we have used for the figures of this article, has
characteristic height $ V_0 $ and characteristic asymmetry $ a =
1/3 $ (where $aL$ is defined as the minimum distance between a
minimum and a maximum of the potential, with $L$ being the period
of the potential).

\bigskip

We have already provided estimations for both the deterministic
part of the net force per particle and the amplitude of its
fluctuations. This will allow us to calculate the average current.
\par

First, we compute the characteristic times during which the potential remains
on, $ \ton $, and off, $ \toff $. In the threshold control
protocol the switching happens when the force crosses the
threshold value with the appropriate slope [see Eq.~\eqref{alfa}].
When the threshold is crossed the equality $u_\nu=f_\nu(t_\nu)$ is satisfied,
with $f_\nu(t)$ the
value of the net force a time $t$ after a switching. Therefore, using the
quasideterministic approximation~\eqref{cuasideterminista} we
obtain for the characteristic times
\be  \label{umbral_periodo}
 |f_\nu^\infty(t_\nu)| - \Sigma = |u_\nu| .
\ee
Using Eq.~\eqref{exponencial} we get the following explicit equations
for the characteristic times
\be
t_\nu = \tau_\nu+\frac{1}{\lambda_\nu}\ln\frac{|C_\nu|}{|u_\nu|+\Sigma},
  \label{tnu}
\ee
with $ \Sigma $ given by Eq.~\eqref{Sigma}. Moreover,
Eq.~\eqref{exponencial} implies that this approximation is valid
for $t_\nu \gtrsim \tau_\nu$, where $\tau_\nu$ are the transient
times for each dynamics (afterwards, Eq.~\eqref{exponencial} is a
good approximation). This implies $|u_\nu|+\Sigma \ll |C_\nu|$,
that can be expressed as $|u_\nu|+\Sigma
\ll\max_t|f_\nu^\infty(t)|$ by using $|C_\nu| \sim
\max_t|f_\nu^\infty(t)|$. As $\Sigma \sim 1/\sqrt N$, we see that
this approximation is valid for small thresholds and large number
of particles.

We now compute the average displacement of the center-of-mass
during an on-off period. Note that the center-of-mass moves only
when the potential is `on', because when it is `off' the dynamics
is purely diffusive. Therefore, as the center-of-mass position is
$ x_{\rm cm} = \sum_i x_i / N $, its average displacement during an on-off
cycle in the many particle case is given by using the evolution
equations~\eqref{langevin} as
\begin{equation}
\xmedia(\ton)=\frac{1}{\gamma}\int_0^{\ton} \fon^\infty(t)\;dt.
\end{equation}
The integration of the late time expression \eqref{exponencial} with
$\nu=\mbox{on}$ suggests a functional form
\begin{equation}\label{fit}
\xmedia(\ton)= \Delta \xon\left( 1-e^{-\ton/{\Delta \ton}}\right).
\end{equation}
This functional form fits well the function $\xmedia(\ton)$
obtained from the numerical integration of the Fokker-Planck
equation, and this fit is used to determine $\Delta \xon$ and
$\Delta \ton$. We have seen that the inclusion of the
characteristic time $\Delta \ton$ improves the analytical results
obtained in Ref.~\cite{prl} (there it was assumed $ \xmedia(\ton)=
\Delta \xon $). This better estimation of the average displacement
improves the results for the intermediate regime of not-so-large
number of particles. Furthermore, the whole expression~\eqref{fit}
is also necessary to improve the results for nonzero thresholds.
When thresholds are enlarged the frequency of switching increases
and therefore the times $\ton$ decrease. This implies a shorter
displacement, as Eq.~\eqref{fit} predicts.
\par

The previous results allows us to give an approximate expression
for the average center-of-mass velocity in the stationary regime,
\begin{equation}\label{vmediaNgrande}
\vmedia\equiv\lim_{t\to \infty}\frac{\xcm(t)-\xcm(0)}{t}= \frac{\Delta
\xon}{\ton+\toff}\left( 1-e^{-\ton/{\Delta \ton}}\right)
= \frac{\Delta\xon \left[1-A \,(\uon+\Sigma)^{{1}/{(\lambdaon \Delta\ton)}}
\right]}
{B-\frac{1}{\lambdaon}\ln(\uon+\Sigma)-\frac{1}{\lambdaoff}
\ln(|\uoff|+\Sigma)},
\end{equation}
with $ \Sigma $ given by Eq.~\eqref{Sigma}, and $ A $ and $ B $ given by
$$
A = e^{-\tauon/\Delta\ton} \Con^{-1/(\lambdaon\Delta\ton)} \;,
    \quad\quad
B = \tauon + \tauoff + \frac{1}{\lambdaon}\ln\Con
    + \frac{1}{\lambdaoff}\ln|\Coff| \;.
$$
The final expression in Eq.~\eqref{vmediaNgrande} shows the explicit dependence
on the thresholds $ \uon $, $ \uoff $, and on the amplitude of the force
fluctuations $ \Sigma $; all the other parameters are determined
by the dynamics for an infinite number of particles with zero thresholds.
Eq.~\eqref{vmediaNgrande} has been obtained in the
quasideterministic approximation and therefore is valid when the
number of particles $ N $ is large and the thresholds are small as
discussed after Eq.~\eqref{tnu}. We have verified that it gives good
estimations inside its regime of validity. In particular, for zero
thresholds Eq.~\eqref{vmediaNgrande} is better than the formula
obtained in Ref.~\cite{prl} thanks to the introduction of the
characteristic time $ \Delta \ton $. (The formula in
Ref.~\cite{prl} is recovered for $ \uon = \uoff = 0 $ and $ \Delta
\ton = 0 $.)

\begin{figure}
\begin{center}
\includegraphics [scale=0.6] {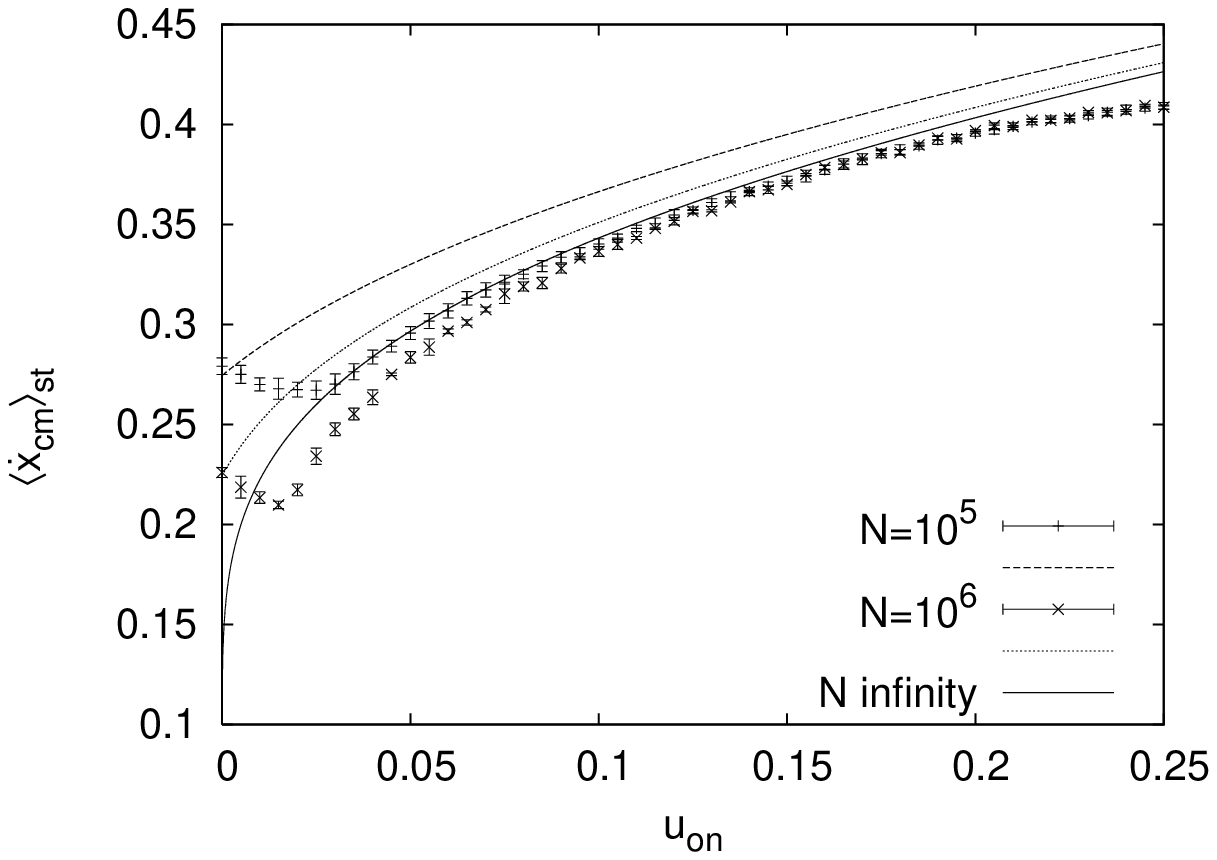}
\end{center}
\caption{Average of the center-of-mass velocity $\vmedia$ as a function of the
  threshold $\uon$ for numbers of particles $N=10^5, 10^6$ and the limit
  $N\to\infty$
  for the potential~\eqref{smoothpot} with $V_0=5
  k_BT$. Analytical quasideterministic approximation~\eqref{vmediaNgrande}
  (lines) and numerical results from Langevin equations~\eqref{langevin}
  (points with error bars). We have taken  $\uoff=-\uon$. (Units: $L=1$,
  $D=1$, $k_BT=1$.)
  }\label{uon_large}
\end{figure}

\begin{figure}
\begin{center}
\includegraphics [scale=0.6] {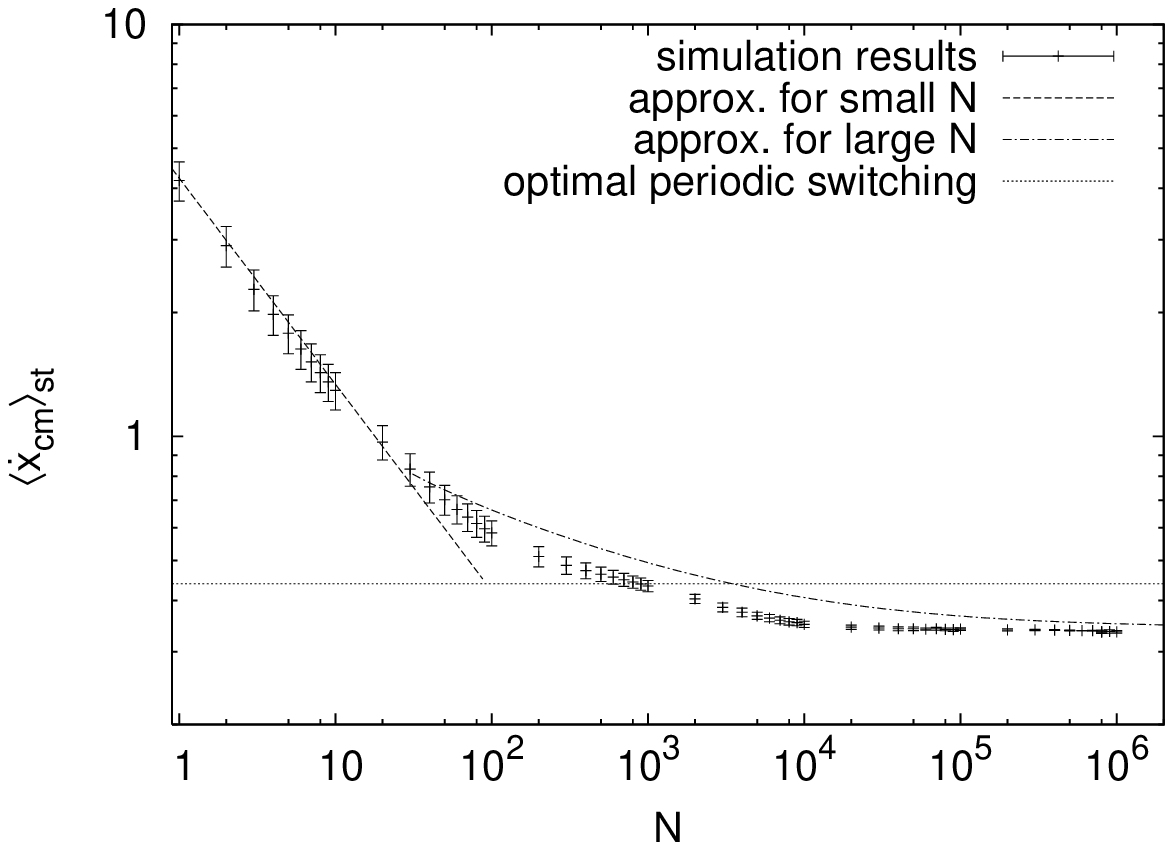}
\end{center}
\caption{Average of the center-of-mass velocity $\vmedia$ as a function of the
  number of particles $N$ for
  the potential~\eqref{smoothpot} with $V_0=5 k_BT$ and for
  thresholds $\uon=0.1$ and $\uoff=-0.1$. The simulations results
  obtained solving numerically the Langevin equations~\eqref{langevin}
  (points with error bars) are compared with the quasideterministic
  approximation for large $N$ [Eq.~\eqref{vmediaNgrande}] and the pure
  stochastic
  approximation for small $N$ [Eq.~\eqref{I1}]. The dotted horizontal straight
  line corresponds to the periodic switching protocol with optimal
  periods. (Units: $L=1$, $D=1$, $k_BT=1$.)}\label{N_01}
\end{figure}

\begin{figure}
\begin{center}
\includegraphics [scale=0.6] {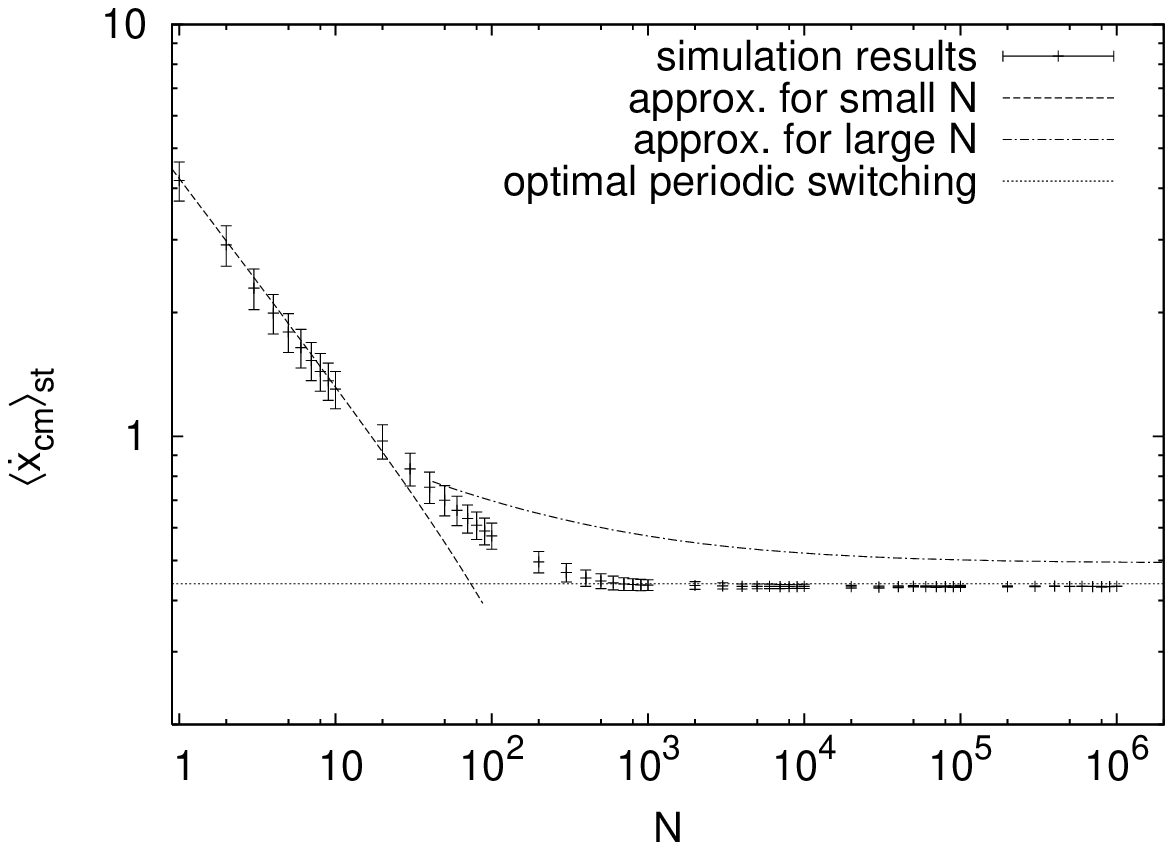}
\end{center}
\caption{Average of the center-of-mass velocity $\vmedia$ as a function of the
  number of particles $N$ for
  the potential~\eqref{smoothpot} with $V_0=5 k_BT$ and for
  thresholds $\uon=0.6$ and $\uoff=-0.4$ (optimal values for
  $N\to\infty$). The simulations results
  obtained solving numerically the Langevin equations~\eqref{langevin}
  (points with error bars) are compared with the quasideterministic
  approximation for large $N$ [Eq.~\eqref{vmediaNgrande}] and the pure
  stochastic
  approximation for small $N$ [Eq.~\eqref{I1}]. The dotted horizontal straight
  line corresponds to the periodic switching protocol with optimal
  periods. (Units: $L=1$, $D=1$, $k_BT=1$.)}\label{N_max}
\end{figure}

Figs.~\ref{uon_large}-\ref{N_max} compare the predictions of the
quasideterministic approximation, Eq.~\eqref{vmediaNgrande}, with
the numerical results for the threshold control protocol applied
with the potential \eqref{smoothpot} and $ V_0 = 5 k_B T $. For
this potential the fit to the Fokker-Planck evolution gives
$\Con=0.67 k_BT/L$, $\tauon=0.058 L^2/D$, $\lambdaon=28 D/L^2$,
$\Coff=-0.74 k_BT/L$, $\tauoff=0.037 L^2/D$, $\lambdaoff=39
D/L^2$, and $\Delta \xon=0.08 L$, $\Delta \ton=0.05 L^2/D$.
\par

In Fig.~\ref{uon_large} we plot the current as a function of the
threshold $\uon$ (with $\uoff=-\uon$) comparing the
quasideterministic approximation~\eqref{vmediaNgrande} and the
numerical results obtained from the Langevin evolution
equations~\eqref{langevin}. We see that the quasideterministic
approximation gives a good estimation of the current. However, it
fails to predict the minimum located at low threshold values.
This minimum is caused by a secondary effect that has not been
accounted in the deduction of the analytic formula. This secondary
effect is due to the fact that nonzero thresholds have the
disadvantage of not being instantly optimal, because they imply
switching on the potential when the force is still negative and
switching off the potential when the force is still positive.
In addition, for very small thresholds the switchings are not induced
much earlier than they would be with zero thresholds due to the
force fluctuations.
Thus, there is a minimum located at thresholds of order
$1/\sqrt{N}$, the magnitude of the force fluctuations. For larger threshold
this secondary effect of the thresholds is overcompensated by the
main effect of avoiding the undesired trapping of the dynamics.
This main effect allows to have similar average displacements of the
particles in a shorter on-off cycle time. Therefore, larger thresholds
increase the average center-of-mass velocity.

\par

Figs.~\ref{N_01} and \ref{N_max} compare analytic and numerical
results for the current as a function of the number of particles
for fixed nonzero thresholds: Fig.~\ref{N_01} for $\uon=-\uoff=0.1
k_BT/L$ and Fig.~\ref{N_max} for $\uon = 0.6 k_BT/L$ and $\uoff
= -0.4 k_BT/L$ (which are the optimal values for an infinite
number of particles). In Fig.~\ref{N_01} we see that the
quasideterministic approximation gives a good estimation for large
number of particles. In Fig.~\ref{N_max} the estimate is more
rough due to the fact that the thresholds do not strictly verify the
validity condition of the quasideterministic approximation ($
|u_\nu| + \Sigma \ll |C_\nu| $). Another interesting result we
have found is that for fixed nonzero thresholds the average
velocity as a function of $N$ tends to a constant asymptotic value
for large number of particles, as Eq.~\eqref{vmediaNgrande} predicts.
For an infinite number of particles the force fluctuations vanishes,
thus, this asymptotic value is given by Eq.~\eqref{vmediaNgrande}
evaluated at $ \Sigma = 0 $.
See Figs.~\ref{N_01} and \ref{N_max}.

The optimal threshold protocol gives the same
current or better than the optimal periodic control~\cite{europhys}
(Fig.~\ref{N_max}). In particular, for an infinite number of
particles the force fluctuations become negligible and the
threshold control becomes equivalent to a periodic switching.
The relation between the thresholds and the periods~\cite{europhys}
\be
u_\nu = f_\nu^\infty({\cal T}_\nu)   \label{u_f}
\ee
is obtained here as the limit $N\to\infty$, i.e. $
\Sigma = 0 $, of Eq.~\eqref{umbral_periodo}.
This relation permits to get the optimal thresholds
for an infinite number of particles from the optimal periods just
using the functions $ \fon^\infty(t) $ and $ \foff^\infty(t) $
obtained numerically from the Fokker-Planck equation. This avoids
the need of integrating numerically $ N $ coupled Langevin
equations for large values of $ N $.
We have numerically checked that the expression~\eqref{u_f}
gives the optimal thresholds (see Sec.~\ref{general} and Fig.~\ref{iso}).

\subsubsection{Few particles: pure stochastic approximation}\label{few}

When we have few particles the situation is the opposite to that
considered in the previous section and the net force has nearly a
pure stochastic behavior. A binomial distribution is found for the
net force probability distribution, $ p(f) $, in Ref.~\cite{prl}
under the approximations that the position of the particles are
statistically independent and that the probability of finding a
particle in the interval $[0,aL]$ is $a$. For simplicity this
binomial distribution for the net force can be approximated by a
Gaussian distribution
\begin{equation}
p(f) \simeq \frac{1}{\sqrt{2\pi\Sigma^2}} \;
e^{-\frac{f^2}{2\Sigma^2}} ,
\end{equation}
with $\Sigma$ the amplitude of the fluctuations of the net force,
that is given by Eq.~\eqref{Sigma}. Neglecting the time
correlations in the net force, the average center-of-mass velocity
for the threshold protocol [Eq.~\eqref{alfa}] is given by
\begin{equation} \label{vmedia_con_integrales}
\vmedia= \frac{1}{\gamma} \int_{\uon}^{\infty} f p(f)\;df +
\frac{1}{\gamma} \int_{\uoff}^{\uon} f p_+(f)\; df \;,
\end{equation}
with $ p_+(f) $ the probability of having a net force $ f $ and a
non-negative value of $ \dfdt $ [$ p_+(f) \sim p(f)/2 $]. This
implies that, in the validity range of this small $N$
approximation [$ \Sigma \gtrsim \max_t |f^\infty(t)|$], the
current is a decreasing function of the threshold $\uon$, as can
be easily proven comparing the results for $ \uon^\prime $ and $
\uon $ with $0\leq \uon^\prime<\uon$.
Eq.~\eqref{vmedia_con_integrales} gives
\begin{equation}
\vmedia(\uon^\prime) - \vmedia(\uon) = \frac{1}{\gamma}
   \int_{\uon^\prime}^{\uon} f p_-(f)\;df,
\end{equation}
with $ p_-(f) \equiv p(f) - p_+(f) \geq 0 $. Thus, the last term
in the previous expression is non-negative implying
\begin{equation}  \label{decreciente}
\vmedia(\uon^\prime)-\vmedia(\uon)\geq 0.
\end{equation}
Analogously, it can be shown that for $ 0 \geq \uoff^\prime >
\uoff $ we have $ \vmedia(\uoff^\prime)-\vmedia(\uoff) =
(1/\gamma)\int_{\uoff}^{\uoff^\prime}(-f) p_+(f) df \geq 0 $. This
shows that the average center-of-mass velocity is a decreasing
function for increasing modulus of the thresholds. Therefore, for
small $N$ we get the maximum current for zero thresholds.

\bigskip

\begin{figure}
\begin{center}
\includegraphics [scale=0.6] {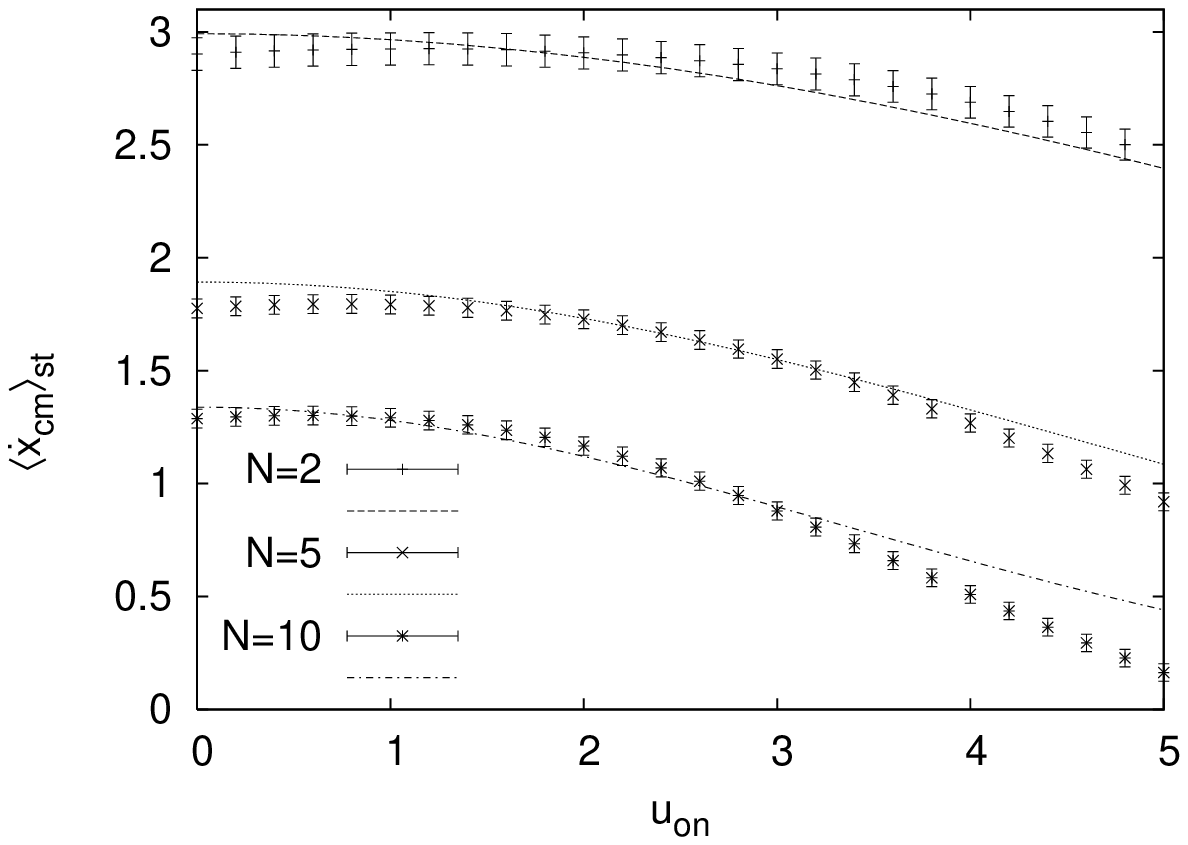}
\end{center}
\caption{Average of the center-of-mass velocity $\vmedia$ as a function of the threshold $\uon$ for $N=2,5,
  \mbox{ and }10$ particles for the potential~\eqref{smoothpot} with $V_0=5
  k_BT$. Analytical pure stochastic approximation~\eqref{I1} (lines) and
  numerical results from Langevin
  equations~\eqref{langevin} (points with error bars) are compared.
  We have taken
  $\uoff=-\uon$. (Units: $L=1$, $D=1$, $k_BT=1$.)}\label{uon_small}
\end{figure}

For small thresholds we have found an approximate explicit
analytical expression for the current. If $ \uoff \simeq - \uon $
the contribution of the second integral in
Eq.~\eqref{vmedia_con_integrales} is generally small, because it
is the integration of a nearly odd function in a nearly symmetric
interval around zero. On the other hand, the contribution of the
first integral is greater provided the thresholds are small
($\uon\lesssim \Sigma$). Then, neglecting the second integral we
obtain
\be  \label{I1}
\vmedia \simeq \frac{\Sigma}{\gamma\sqrt{2\pi}} \; e^{-\frac{\uon^2}{2\Sigma^2}}.
\ee
(Note that for $ \uon = 0 $ we recover the zero threshold result
found in Ref.~\cite{prl}.) This expression, Eq.~\eqref{I1}, gives
good predictions when we have few particles and small thresholds.
In particular, we show in Figs.~\ref{N_01}-\ref{uon_small} that it
correctly predicts the threshold and particle number dependence of
the current, even for $ \uon \sim \Sigma \simeq 3.4 $ when $ N =
10 $ (Fig.~\ref{uon_small}).

\subsection{General thresholds}\label{general}

In the previous subsection we have studied the threshold protocol
when the moduli of the thresholds are small, obtaining
approximate analytical expressions for the current. In contrast,
in this subsection we study the threshold protocol for general
thresholds (that are in general beyond the applicability range of
the previous analytical expressions). This study is done
performing numerical simulations of the Langevin equation of the
threshold protocol for general values of the thresholds.

\subsubsection{$\uoff=-\uon$}

Let us discuss first the results for thresholds that are related
by $\uoff=-\uon$.

In the few particle case, when the thresholds are small the current
decreases exponentially with the square of the threshold as we
have already seen [see Eq.~\eqref{I1}]. However, as the rate of
the exponential is small, we nearly have a plateau near the
maximum at zero thresholds, as shown in Figs.~\ref{uon_small} and
\ref{uon}. On the other hand, for very large thresholds
Eq.~\eqref{I1} is no longer valid and the current decreases faster
than the exponential. Note that the current continues to be a
decreasing function, as predicted by Eq.~\eqref{decreciente}
(valid for any threshold values in the few particle case). See
Figs.~\ref{uon_small} and \ref{uon}.

In contrast, in the many particle case the maximum of the current
is no longer at zero thresholds, but at a finite value. As we have
explained before, the introduction of thresholds has the advantage
of inducing earlier switchings. This avoids the undesired trapping
that otherwise is present for large $N$ implying low current
values. The presence of thresholds allows to have similar average
displacements of the particles in a shorter on-off cycle time, and
therefore increases the average center-of-mass velocity. However,
if the thresholds are too large the losses in the displacement
become more important than the gains of shortening the on-off
cycle time. Therefore, the current has a maximum located at a
finite value of the thresholds in the many particle case
(Fig.~\ref{uon}). (The tiny minimum in the small threshold  region
is related to another effect: the disadvantages of choosing a not
instantly optimal protocol. For a more detailed explanation see
Subsec.~\ref{many}.)
\newline
Another important result in the many particle case is that the maximum
obtained for the current as a function of the threshold magnitude
is quite flat and nearly independent of the number of particles.
See Fig.~\ref{uon}.

In summary, in the many particle case the current has a maximum
for nonzero thresholds whose position is nearly independent of the
number of particles. On the other hand, for few particles the
current is maximum for zero thresholds. However, in the few
particle case the current is nearly the same up to thresholds of
the order of the thresholds that give the maximum for the many
particle case (see Figs.~\ref{uon_small} and \ref{uon}). This has an important
implication: the optimal thresholds values for the many particle
case give currents close to the maximum for \emph{any} number of
particles.\par

\subsubsection{$ \uoff \neq -\uon $}

\begin{figure}
\begin{center}
\includegraphics [scale=0.6] {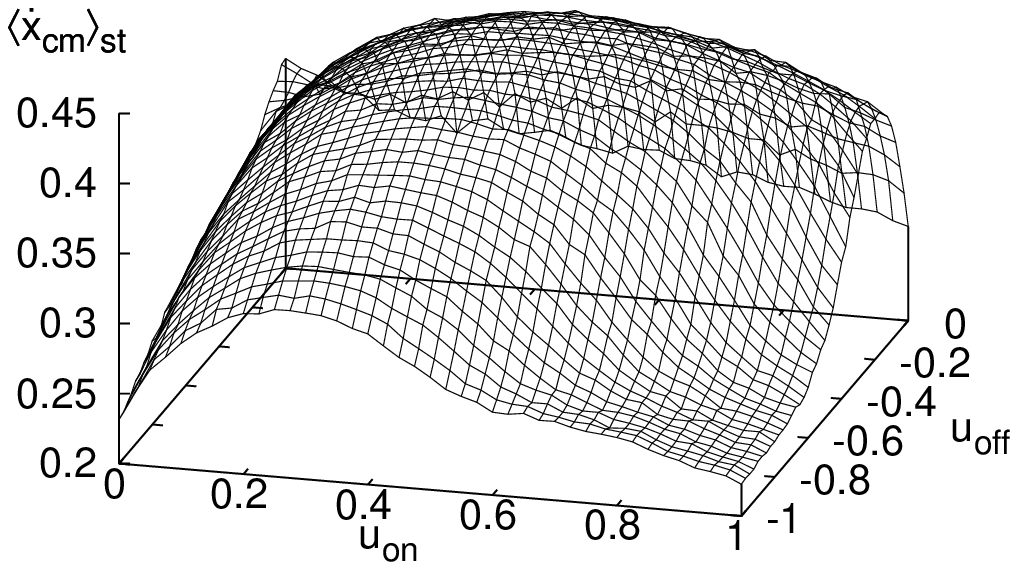}
\end{center}
\caption{Thresholds dependence of the average of the
  center-of-mass velocity $\vmedia$ for $N=10^4$ particles in the
  potential~\eqref{smoothpot} with
  $V_0=5 k_BT$. The grid has been obtained integrating numerically Langevin
  equations~\eqref{langevin} for different thresholds $\uon$  and $\uoff$
  (Units: $L=1$, $D=1$, $k_BT=1$.)}
\label{uon_uoff}
\end{figure}

\begin{figure}
\begin{center}
\includegraphics [scale=0.6] {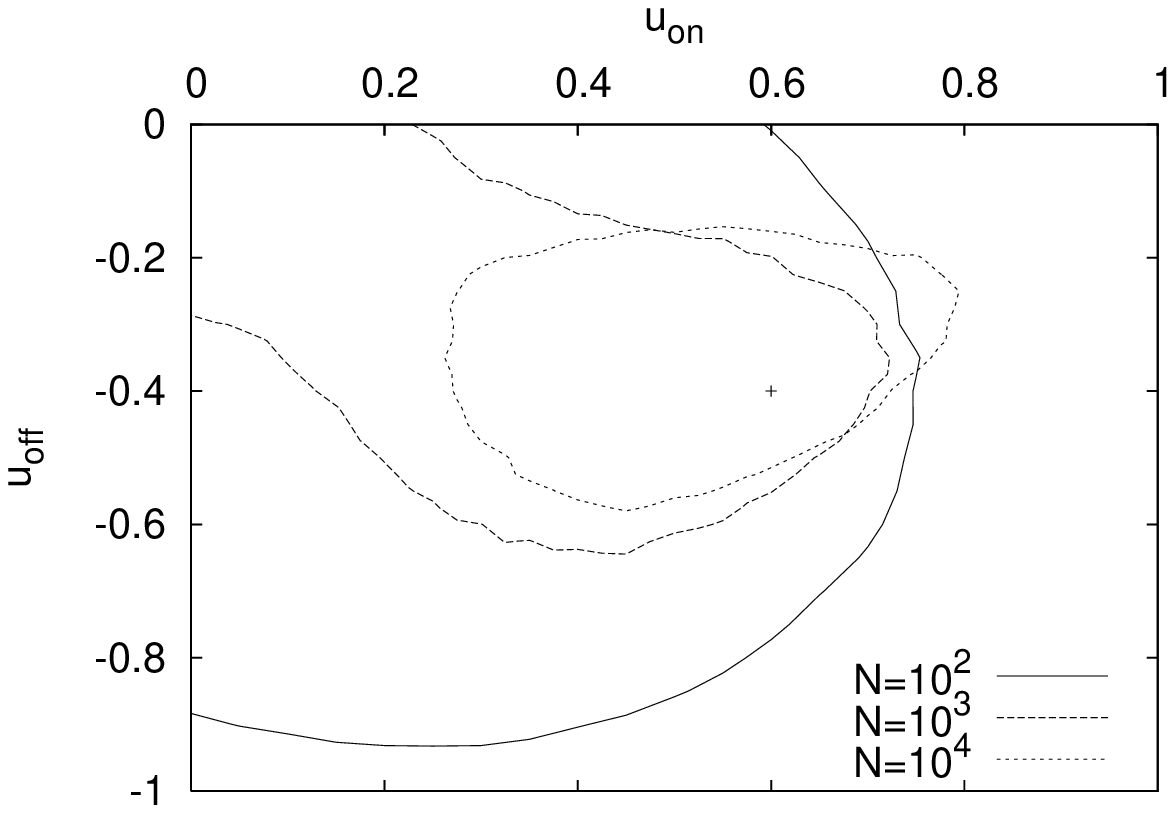}
\end{center}
\caption{Thresholds contour lines corresponding to the value of the average of
  the center-of-mass velocity $\vmedia$ $5\%$ below its maximum for $N=10^2$,
  $N=10^3$ and $N=10^4$ particles in the potential~\eqref{smoothpot} with
  $V_0=5 k_BT$. The contour line for $N=10^5$ is already very similar to that
  for   $N=10^4$. The point corresponds to the optimal thresholds for $N\to
  \infty$ obtained from the optimal periods using Eq.~\eqref{u_f}.
  (Units: $L=1$, $D=1$, $k_BT=1$.)}
  \label{iso}
\end{figure}

The study of the current for completely general
thresholds $ \uon $ and $ \uoff $ (without restrictions) reveals
that the behavior is analogous to that described previously. In
fact, the optimal thresholds for large number of particles are
located not far from the line $ \uon = -\uoff $, and these thresholds
give currents close to the maximum for any number of particles. (See
Figs.~\ref{uon_uoff} and \ref{iso}.)\par

\begin{figure}
\begin{center}
\includegraphics [scale=0.6] {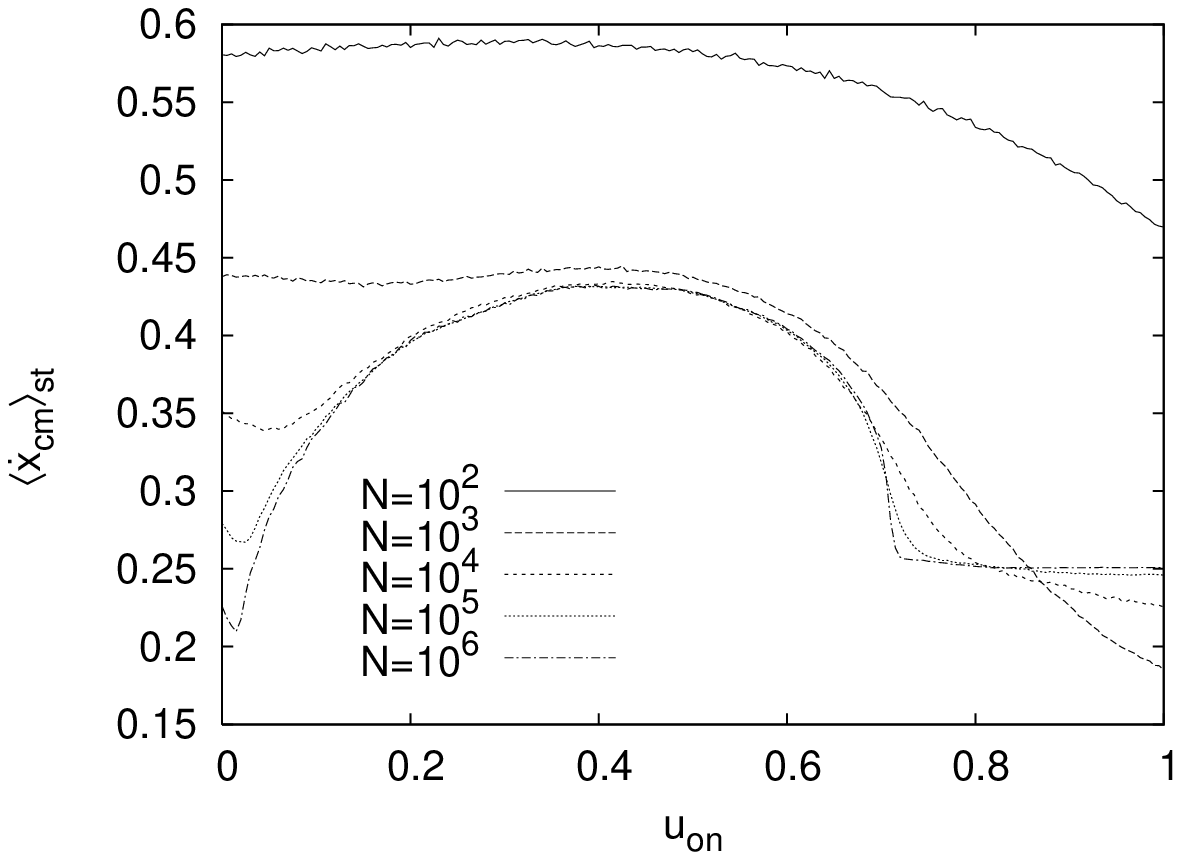}
\end{center}
\caption{Average of the center-of-mass velocity $\vmedia$ as a function of the
  threshold $\uon$ with $\uoff=-\uon$ for various $N$. The lines correspond to
  the numerical solution of the Langevin equations~\eqref{langevin} for the
  potential~\eqref{smoothpot} with $V_0=5 k_BT$. (Units: $L=1$, $D=1$,
  $k_BT=1$.)}\label{uon}
\end{figure}

As we have already commented in the previous section, for an
infinite number of particles the force fluctuations becomes
negligible and the threshold protocol becomes equivalent to a
periodic switching. This implies the relation~\eqref{u_f} between
the optimal periods $\calTon$ and $ \calToff$, and the optimal
thresholds $ \uon $ and $ \uoff $, that we have numerically checked
(see Fig.~\ref{iso}). Therefore, these relations
permit to obtain the optimal thresholds for an infinite number of
particles from the optimal periods, just using the functions $
\fon^\infty(t) $ and $ \foff^\infty(t) $ obtained numerically from
the Fokker-Planck equation. These thresholds give good results for
large number of particles. Moreover, it is important to note that
these thresholds values also give currents close to the maximum
in the few particles case due to the smooth dependence for small
thresholds (see Figs. \ref{uon}-\ref{iso}).\par

In particular, we have seen that for the potential
\eqref{smoothpot} with $ V_0 = 5 k_B T $ the optimal switching
periods are approximately $ \calTon=0.06 L^2/D $ and $
\calToff=0.05 L^2/D$. Therefore, with just a Fokker-Planck
simulation for the potential we have found that a good estimation
of the optimal thresholds is given by $\uon=\fon^\infty
(\calTon)=0.6 k_BT/L$ and $\uoff=\foff^\infty (\calToff)=-0.4
k_BT/L$, in good agreement with Fig.~\ref{iso}.

\section{Conclusions}\label{conclusions}

In this paper we have analyzed the \emph{threshold control}
protocol for a collective flashing ratchet. We have studied the
threshold dependence of the current in this closed-loop control
protocol. The quasideterministic (for many particles)
approximation \cite{prl} has been improved through the
introduction of an additional characteristic time giving better
results for not-so-many particles. Both the quasideterministic and
the stochastic (for few particles) approximations \cite{prl} have
been applied to the threshold control protocol. This has led to
analytical expressions for large and small number of particles. We
have computed numerically the current dependence on the thresholds
and on the number of particles obtaining a good agreement between
analytical and numerical results in the validity range of our
assumptions. We have also compared these results with the
\emph{optimal periodic switching} protocol.\par

We have seen that for many particles the current has a maximum for
nonzero thresholds whose position is nearly independent of the
number of particles. On the other hand, for few particles we have demonstrated
that the current increases as thresholds moduli decrease, so the maximum
current is reached at zero thresholds. However, the current is
nearly the same up to thresholds of the order of the optimal
thresholds for the many particle case. This implies that the
optimal thresholds values for the many particle case give currents
close to the maximum for any number of particles. The optimal thresholds for
an infinite number of particles can be obtained from the optimal periods of
the periodic protocol just solving the Fokker-Planck
equation in two particular cases (potential `on' and `off',
see Section~\ref{small}). Therefore, we
can get a good estimation of the optimal thresholds for many
particles, that also gives currents close to the optimal for any
number of particles as we have shown.\par

The closed-loop threshold control gives the same
current as the optimal protocols for the one particle case and for
an infinite number of particles, and it gives high currents in
between. However, obtaining the best protocol
for the maximization of the current is still an open question.

In this work, and in previous ones~\cite{prl,europhys}, we have
seen that thanks to the information about the fluctuations
obtained through the feedback, the performance of the system can
be increased. This increase of the performance has thermodynamical
limitations that have been studied in a general context for the
efficiency~\cite{hugo}. We are now working in order to get a
deeper understanding of this interplay between the information and
the increase of the performance~\cite{future}.

\acknowledgments

We acknowledge financial support from the Ministerio de Ciencia y
Tecnolog\'{\i}a (Spain) through the Research Projects
BFM2003-02547/FISI and FIS2006-05895. In addition, MF thanks the
Universidad Complutense de Madrid (Spain) and FJC thanks ESF
Programme STOCHDYN for their financial support.

\end{document}